\documentclass[article,preprint,groupedaddress]{revtex4}
\usepackage{epsfig}
\usepackage{graphicx}

\begin{document}
\title{Survival probabilities in the Sisyphus random walk model with absorbing traps}
\author{Shahar Hod}
\affiliation{The Ruppin Academic Center, Emeq Hefer 40250, Israel}
\affiliation{ }
\affiliation{The Hadassah Institute, Jerusalem 91010, Israel}
\date{\today}

\begin{abstract}
\ \ \ We analyze the dynamics of the Sisyphus random walk model, a
discrete Markov chain in which the walkers may randomly return to
their initial position $x_0$. In particular, we present a remarkably
compact derivation of the time-dependent survival probability
function $S(t;x_0)$ which characterizes the random walkers in the
presence of an absorbing trap at the origin. The survival
probabilities are expressed in a compact mathematical form in terms
of the $x_0$-generalized Fibonacci-like numbers $G^{(x_0)}_t$.
Interestingly, it is proved that, as opposed to the standard random
walk model in which the survival probabilities depend linearly on
the initial distance $x_0$ of the walkers from the trap and decay
asymptotically as an inverse power of the time, in the Sisyphus
random walk model the asymptotic survival probabilities decay
exponentially in time and are characterized by a non-trivial
(non-linear) dependence on the initial gap $x_0$ from the absorbing
trap. We use the analytically derived results in order to analyze
the underlying dynamics of the `survival-game', a highly risky
investment strategy in which non-absorbed agents receive a reward
$P(t)$ which gradually increases in time.
\end{abstract}
\bigskip
\maketitle

\section{Introduction}

Random walk is certainly one of the most intriguing concepts in
statistical physics \cite{RW1,RW2,RW3,RW4,RW5,RW6}. In particular,
the dynamics of random walkers in the presence of absorbing traps
have attracted the attention of physicists and mathematicians in the
last three decades. These phenomenology rich models find
applications in diverse areas of physics, such as granular
segregation \cite{DV1}, complex adaptive systems \cite{DV2}, polymer
adsorption \cite{DV3}, the dynamics of epidemic spreading
\cite{DV4}, and the theory of absorbing-state phase transitions
\cite{DV5}. Interestingly, these physical models are characterized
by an asymptotic inverse power-law decay in time of the survival
probabilities of the random walkers \cite{DV2,Notesur}:
\begin{equation}\label{Eq1}
S(t;x_0)=x_0 \sqrt{{{2}\over{\pi}}}t^{-{1\over2}}\  ,
\end{equation}
where $x_0$ is the initial gap between the walkers and the absorbing
trap.

In this paper we shall study an interesting variant of the standard
random walk model, known as the Sisyphus random walk \cite{Sisy}. In
this physical model, which belongs to the family of random walks
with restart mechanisms \cite{RS1,RS2,RS3,RS4,RS5}, the random
walker may move at each time tick one step towards the absorbing
trap (which, we assume, is located at the origin
$x_{\text{trap}}=0$) or, alternatively, may randomly increase her
distance from the absorbing boundary by returning all the way to her
initial position $x_0>0$.

The main goal of the present paper is to analyze the influence of
this restart mechanism on the time-dependent survival probabilities
of the Sisyphus random walkers in the presence of an absorbing trap.
In particular, we here raise the following physically interesting
question: Which type of a random walker has the larger survival
probability? The standard random walker [who, at each time tick, may
take a random step (with probability $1/2$) in the opposite
direction of the absorbing trap and who has no upper bound on her
distance from the trap] or the Sisyphus random walker (who, at each
time tick, may randomly return all the way to her initial position
$x_0$, which also acts as an upper bound on the distance of the
walker from the absorbing trap)?

As we shall explicitly show below, this is a non-trivial question
since the restart mechanism, which characterizes the Sisyphus random
walk model, introduces into the physical system two competing
factors which have {\it opposite} effects on the survival
probabilities of the walkers:
\newline
(1) On the one hand, the presence of an upper bound $x(t)\leq x_0$
on the locations of the Sisyphus walkers prevents them from reaching
the asymptotically safe region $x\to\infty$, thus decreasing the
expected survival probability of these walkers.
\newline
(2) On the other hand, the restart mechanism allows random walkers
in the dangerous zone $x\sim x_{\text{trap}}=0$ to abruptly jump
(within a single time step) into the safer zone $x=x_0$, thus
increasing the survival probability of these lucky walkers.

In this paper we shall use analytical techniques in order to prove
that the restart mechanism in the Sisyphus random walk model with an
absorbing trap {\it decreases} the survival probabilities of the
walkers as compared to the corresponding survival probabilities of
standard random walkers. In particular, we shall explicitly show
below that, instead of the inverse power-law decay (\ref{Eq1}) which
characterizes the survival probabilities of standard random walkers,
the Sisyphus random walkers are characterized by exponentially
decaying asymptotic survival probabilities.

\section{Description of the system}

We consider a discrete time ($t=0,1,2,...$) random walk on the
non-negative integers with an absorbing trap at the origin. The
initial location of the walker is denoted by $x_0>0$. At each
discrete time step the walker randomly (with probability $1/2$) goes
one step to the left (towards the absorbing trap at
$x_{\text{trap}}=0$) or returns to her initial position $x_0$.

We denote by $N_x(t)$ the number of random walkers which are located
at a given position $x$ [with $x\in\{0,1,2,...\}$] at the discrete
time step $t$. It is convenient to use the normalization
\begin{equation}\label{Eq2}
N_{x_0}(t=0)=1\ \ \ \ ; \ \ \ \ N_x(t=0)=0 \ \ \ {\text{for}}\ \ \
x\neq x_0\ .
\end{equation}
In addition, the boundary conditions of the system are given by
\begin{equation}\label{Eq3}
N_{\text{tot}}(t)=1 \ \ \ \ \text{for}\ \ \ \ t=0,1,...,x_0-1\  ,
\end{equation}
where
\begin{equation}\label{Eq4}
N_{\text{tot}}(t)\equiv \sum_{k=1}^{k=x_0}N_k(t)\
\end{equation}
is the total number of untrapped walkers at time $t$.

\section{Survival probabilities in the Sisyphus random walk model}

In the present section we shall present a remarkably compact
derivation of the time-dependent survival probability function
$S(t;x_0)$ which characterizes the Sisyphus random walk model with a
trap, that is the probability that a random walker who is initially
located at $x_0$ has not been absorbed at the origin until the
discrete time $t$ [Taking cognizance of the normalization
(\ref{Eq2}), one realizes that the discrete function
$N_{\text{tot}}(t)$ is the survival probability of the random
walkers at time $t$].

We first note that, by definition, the number of random walkers
$d(t)$ which are absorbed at the origin at the discrete time step
$t$ is given by
\begin{equation}\label{Eq5}
d(t)=N_{\text{tot}}(t-1)-N_{\text{tot}}(t)\  .
\end{equation}
Substituting into (\ref{Eq5}) the relations
\begin{equation}\label{Eq6}
d(t)={1\over2}N_{x=1}(t-1)\  ,
\end{equation}
\begin{equation}\label{Eq7}
N_{x=1}(t-1)=\Big({1\over2}\Big)^{x_0-1}N_{x_0}(t-x_0)\  ,
\end{equation}
and
\begin{equation}\label{Eq8}
N_{x_0}(t-x_0)={1\over2}N_{\text{tot}}(t-x_0-1)\  ,
\end{equation}
[Here we have used the fact that, at each discrete time step, each
random walker has a probability $1/2$ to go one step to the left
(towards the absorbing trap) or to go back to her initial location
$x_0$] one finds the recurrence relation
\begin{equation}\label{Eq9}
N_{\text{tot}}(t)=N_{\text{tot}}(t-1)-\Big({1\over2}\Big)^{x_0+1}N_{\text{tot}}(t-x_0-1)\
\end{equation}
for the number of untrapped random walkers.

It proves useful to define
\begin{equation}\label{Eq10}
N_{\text{tot}}(t)={{f(t;x_0)}\over{2^t}}\  ,
\end{equation}
in which case one obtains from (\ref{Eq9}) the relation
\begin{equation}\label{Eq11}
{{f(t;x_0)}\over{2^t}}={{f(t-1;x_0)}\over{2^{t-1}}}-\Big({1\over2}\Big)^{x_0+1}{{f(t-x_0-1;x_0)}
\over{2^{t-x_0-1}}}\  .
\end{equation}
From Eq. (\ref{Eq11}) one finds that the function $f(t;x_0)$
satisfies the recurrence relation
\begin{equation}\label{Eq12}
f(t;x_0)=2f(t-1;x_0)-f(t-x_0-1;x_0)\  .
\end{equation}
Substituting Eq. (\ref{Eq12}) into itself, one finds the relation
\begin{equation}\label{Eq13}
f(t;x_0)=f(t-1;x_0)+2f(t-2;x_0)-f(t-x_0-2;x_0)-f(t-x_0-1;x_0)\  .
\end{equation}
Repeating this substitution procedure, one finds
\begin{equation}\label{Eq14}
f(t;x_0)=f(t-1;x_0)+f(t-2;x_0)+2f(t-3;x_0)-f(t-x_0-3;x_0)-f(t-x_0-2;x_0)-f(t-x_0-1;x_0)\
.
\end{equation}
In general, repeating this substitution procedure $x_0$ times, one
finally obtains the compact $x_0$-th order recurrence relation
\begin{equation}\label{Eq15}
f(t;x_0)=\sum_{k=1}^{k=x_0}f(t-k;x_0)\ .
\end{equation}

Interestingly, we therefore find from Eq. (\ref{Eq15}) that the
discrete function $f(t;x_0)$ in the survival probability function
(\ref{Eq10}) is the $x_0$-generalized Fibonacci-like sequence, which
is usually denoted in the literature by $G^{(x_0)}_t$ \cite{Fib}.
Taking cognizance of Eqs. (\ref{Eq3}) and (\ref{Eq10}), one finds
the simple boundary conditions
\begin{equation}\label{Eq16}
f(t;x_0)=2^t \ \ \ \text{for}\ \ \ 0\leq t\leq x_0-1\
\end{equation}
and
\begin{equation}\label{Eq17}
f(t=x_0;x_0)=2^{x_0}-1\ .
\end{equation}
The time-dependent function $f(t;x_0)$ in the survival probability
function (\ref{Eq10}) can be expressed in the form \cite{Fib}
\begin{equation}\label{Eq18}
f(t;x_0)=2\sum_{i=1}^{x_0}A(i;x_0)\epsilon^{t-2}_{i}+\sum_{m=0}^{x_0-3}
\Big(2^{m+1}\sum_{j=0}^{m+1}\sum_{i=1}^{x_0}A(i;x_0)\epsilon^{t-2-j}_{i}\Big)+
(2^{x_0}-1)\sum_{i=1}^{x_0}A(i;x_0)\epsilon^{t-1}_{i}\  ,
\end{equation}
where
\begin{equation}\label{Eq19}
A(i;x_0)=(\epsilon_i-1)[2+(x_0+1)(\epsilon_i-2)]^{-1}\
\end{equation}
and $\{\epsilon_1,\epsilon_2,...,\epsilon_{x_0}\}$ are the
characteristic roots of the polynomial equation
\begin{equation}\label{Eq20}
\epsilon^{x_0}-\epsilon^{x_0-1}-\cdots-1=0\  .
\end{equation}

The asymptotic large-$t$ behavior of the function $f(t;x_0)$ can be
obtained by substituting the relation
\begin{equation}\label{Eq21}
f(t;x_0)=\alpha\cdot\beta^t\
\end{equation}
into Eq. (\ref{Eq12}), in which case one finds that the parameter
$\beta$ is determined by the largest positive root of the polynomial
equation
\begin{equation}\label{Eq22}
\beta^{x_0+1}-2\beta^{x_0}+1=0\  .
\end{equation}
For $x_0=\{1,2,3,4,5,...\}$ one finds the corresponding
monotonically increasing values
$\beta(x_0)=\{1,1.618,1.839,1.928,1.966,...\}$. In particular,
assuming $x_0\gg1$ for the initial gap between the Sisyphus random
walkers and the absorbing trap and substituting
\begin{equation}\label{Eq23}
\beta(x_0)=2[1-c(x_0)]\ \ \ \text{with}\ \ \ c\ll1\  ,
\end{equation}
one obtains from (\ref{Eq22}) the relation
\begin{equation}\label{Eq24}
\beta(x_0)=2[1-2^{-(x_0+1)}]\cdot\{1+O[2^{-2(x_0+1)}x_0]\}\
\end{equation}
for the largest positive root of Eq. (\ref{Eq22}). The coefficient
$\alpha$ in (\ref{Eq21}) can be determined by the boundary condition
$f(t=x_0;x_0)=2^{x_0}-1$ [see Eq. (\ref{Eq17})], which yields
\begin{equation}\label{Eq25}
\alpha(x_0)={{1-2^{-x_0}}\over{[1-2^{-(x_0+1)}]^{x_0}}}\  .
\end{equation}

Taking cognizance of Eqs. (\ref{Eq10}), (\ref{Eq21}), and
(\ref{Eq24}), one finds that the time-dependent survival
probabilities of the Sisyphus random walkers in the presence of an
absorbing trap are characterized by the simple asymptotic ratio
\begin{equation}\label{Eq26}
{\cal R^{\text{asym}}}\equiv
{{S(t+1;x_0)}\over{S(t;x_0)}}=1-2^{-(x_0+1)}\ \ \ \ \text{for}\ \ \
\ x_0,t\gg 1\  .
\end{equation}

\section{Numerical confirmation}

It is of physical interest to confirm the validity of the
analytically derived relation (\ref{Eq26}) which characterizes the
late-time survival probabilities of the Sisyphus random walkers in
the presence of an absorbing trap. In Table \ref{Table1} we present
the dimensionless ratio ${\cal R}^{\text{exact}}(t)\equiv
S(t+1;x_0=6)/S(t;x_0=6)$ of the survival probabilities as obtained
numerically from the exact solution (\ref{Eq10}) with the recurrence
relation (\ref{Eq12}). From the data presented in Table \ref{Table1}
one learns that the analytically derived relation (\ref{Eq26}),
which characterizes the asymptotic ($x_0,t\gg 1$) survival
probabilities of the random walkers, agrees remarkably well with the
corresponding ratio ${\cal R}^{\text{exact}}(t)$ as computed
numerically from the exact time-dependent survival probabilities
(\ref{Eq10}).

\begin{table}[htbp]
\centering
\begin{tabular}{|c|c|c|c|c|c|c|c|}
\hline \text{Time step $t$} & \ 7\ \ & \ 10\ \ & \
13\ \ & \ 16\ \ & \ 19\ \ & \ 22 \\
\hline \ ${\cal R}^{\text{exact}}(t)\equiv S(t+1)/S(t)$\ \ \ &\ \
0.992000\ \ \ &\ \ 0.991803\ \ \ &\ \ 0.991795\ \ \ &\ \ 0.991791\
\ \ &\ \ 0.991791\ \ \ &\ \ 0.991791\ \ \\
\hline
\end{tabular}
\caption{Survival probabilities in the Sisyphus random walk model
with a trap at the origin. We display the time-dependent
dimensionless ratio ${\cal R}^{\text{exact}}(t)\equiv
S(t+1;x_0)/S(t;x_0)$ of the survival probabilities as obtained from
the exact solution (\ref{Eq10}) with the recurrence relation
(\ref{Eq12}). The data presented is for the case $x_0=6$. One finds
that the exact (numerically computed) survival probabilities of the
Sisyphus random walkers are characterized with a high degree of
accuracy by the analytically derived asymptotic relation ${\cal
R}^{\text{asym}}=1-2^{-7}\simeq 0.992188$ [see Eq. (\ref{Eq26})].}
\label{Table1}
\end{table}

\section{The survival game}

In the present section we shall use the analytically derived results
of the previous sections in order to analyze the underlying dynamics
of the `survival-game'. In this popular game, which may mimic highly
risky investment strategies, the agent (the random walker) declares,
at the beginning of the game, the number $t$ (with $t\geq x_0$) of
time steps (bets) that she intends to take during the game. The
rules of the survival game are very simple: if the agent has not
been absorbed during the $t$-step game (that is, the agent has not
made $x_0$ failed bets in a row during the game), then she receives
a reward, where the prize function $P(t)$ is a monotonically
increasing function of the survival time $t$. On the other hand,
absorbed agents (agents who have made $x_0$ failed bets in a row)
are doomed to lose all their resources (money).

It is interesting to ask: Given a monotonically increasing
time-dependent reward function $P(t)$ for the lucky (non-absorbed)
agents, what time $t_{\text{opt}}$ is statistically the best
(optimal) time to quit this highly risky game? The answer to this
practical question is given, for a given value of the initial
distance $x_0$ of the agent from the absorbing trap [note that, in
the survival game, the parameter $x_0-1$ denotes the maximal number
of failed bets in a row which are allowed for the agent (the random
walker) before she is forced to quit this highly risky game (that
is, before she loses all her resources)] by the extremum point of
the combined reward function
\begin{equation}\label{Eq27}
R(t;x_0)\equiv S(t;x_0)\times P(t)\  .
\end{equation}
In particular, taking cognizance of Eqs. (\ref{Eq10}) and
(\ref{Eq21}), the equation
\begin{equation}\label{Eq28}
{{d[S(t;x_0)P(t)]}\over{dt}}=0
\end{equation}
yields the characteristic functional expression
\begin{equation}\label{Eq29}
\ln[\beta(x_0)/2]=-{{(dP/dt)_{t=t_{\text{opt}}(x_0)}}\over{P[t=t_{\text{opt}}(x_0)]}}\
\ \ ; \ \ \ t_{\text{opt}}\gg1\
\end{equation}
for the statistically optimal asymptotic quitting time.

If the time-dependent reward function $P(t)$ grows slower than an
exponent, then our analysis reveals the fact that there is an
optimal asymptotic quitting time [see Eq. (\ref{Eq29})] in the
survival game. For example, if the monotonically increasing reward
function is described by a simple power-law of the form
\begin{equation}\label{Eq30}
P(t)=At^{\gamma}\ \ \ ; \ \ \ \gamma>0\  ,
\end{equation}
then the optimal asymptotic quitting time in the survival game is
given by the simple expression
\begin{equation}\label{Eq31}
t_{\text{opt}}(x_0)=-{{\gamma}\over{\ln[\beta(x_0)/2]}}\  .
\end{equation}
In the $x_0\gg1$ regime one can use the relation
\begin{equation}\label{Eq32}
\ln(1-z)=-z+O(z^2)\ \ \ \text{for}\ \ \ z\ll1
\end{equation}
in order to express the optimal asymptotic quitting time
(\ref{Eq31}) in the compact form [see Eq. (\ref{Eq24})]
\begin{equation}\label{Eq33}
t_{\text{opt}}(x_0)=\gamma\cdot 2^{x_0+1}\  .
\end{equation}
The corresponding optimal asymptotic reward $R(t_{\text{opt}};x_0)$
in the survival game is given by [see Eqs. (\ref{Eq10}),
(\ref{Eq21}), (\ref{Eq27}), (\ref{Eq30}), and (\ref{Eq33})]
\begin{equation}\label{Eq34}
R(t_{\text{opt}};x_0)=A[t_{\text{opt}}(x_0)]^{\gamma}\alpha(x_0)[\beta(x_0)/2]^{t_{\text{opt}}(x_0)}\
.
\end{equation}
Taking cognizance of Eqs. (\ref{Eq24}), (\ref{Eq25}), and
(\ref{Eq33}), one can express (\ref{Eq34}) in the explicit
$x_0$-dependent form
\begin{equation}\label{Eq35}
R(t_{\text{opt}};x_0)=A{{1-2^{-x_0}}\over{[1-2^{-(x_0+1)}]^{x_0}}}
\Big\{\gamma 2^{x_0+1}[1-2^{-(x_0+1)}]^{2^{x_0+1}}\Big\}^{\gamma}\ .
\end{equation}

\section{Summary and Discussion}

We have analyzed the dynamics of the Sisyphus random walk model, an
interesting variant of the standard random walk model in which the
walkers may randomly return to their initial position $x_0>0$. In
particular, a remarkably compact derivation of the characteristic
survival probability function $S(t;x_0)$ of the random walkers in
the presence of an absorbing trap at the origin has been presented.
Intriguingly, we have shown that the time-dependent survival
probabilities can be expressed in a compact mathematical form in
terms of the $x_0$-generalized Fibonacci-like numbers [see Eqs.
(\ref{Eq10}) and (\ref{Eq15})]
\begin{equation}\label{Eq36}
S(t;x_0)={{G^{(x_0)}_t}\over{2^t}}\  .
\end{equation}

Interestingly, we have explicitly proved that the survival
probabilities in the Sisyphus random walk model with an absorbing
trap are drastically different from the corresponding survival
probabilities of the standard random walk model. In particular, we
have shown that:
\newline
(1) As opposed to the standard random walk model in which the
survival probabilities decay asymptotically as an inverse power of
time [see Eq. (\ref{Eq1})], in the Sisyphus random walk model the
asymptotic survival probabilities decay exponentially in time [see
Eq. (\ref{Eq26})].
\newline
(2) As opposed to the standard random walk model in which the
asymptotic survival probabilities depend linearly on the initial
position $x_0$ of the walkers [see Eq. (\ref{Eq1})], in the Sisyphus
random walk model the survival probabilities are characterized by a
non-trivial (non-linear) dependence on the initial gap $x_0$ between
the walkers and the absorbing trap [see Eqs. (\ref{Eq25}) and
(\ref{Eq26})].

\bigskip
\noindent
{\bf ACKNOWLEDGMENTS}
\bigskip

This research is supported by the Carmel Science Foundation. I would
like to thank Yael Oren, Arbel M. Ongo, Ayelet B. Lata, and Alona B.
Tea for helpful discussions.


\begin{thebibliography}{99}
\bibitem{Notesur} The time-dependent survival probability function $S(t;x_0)$
is the probability that a random walker, who is initially located at
$x_0$, has not been absorbed at the origin until the discrete time
$t$.
\end{thebibliography}
\end{document}